\documentclass[12pt]{article}

\usepackage[margin=1in]{geometry}

\usepackage{natbib}

\newcommand{\md}{\mathrm{d}}

\newcommand{\mE}{\mathbb{E}}

 \newcommand{\Cov}{\mbox{Cov}}
 \newcommand{\bitem}{\begin{itemize}}
 \newcommand{\eitem}{\end{itemize}}

 \newcommand{\hf}{{ \widehat {f} }}

\def\pr{\mathbb{P}}

\def\I{\mathbb{I}} %

\newcommand{\hde}{\wh{\delta}}

\newcommand{\tf}{\widetilde{f}}

\usepackage[figuresright]{rotating}

\usepackage{amsmath}
\usepackage{graphicx} %
\usepackage{hyperref}
\usepackage{amsfonts,mathrsfs,amssymb%
}
\usepackage{footnote}

\usepackage{subfigure}
\usepackage{multirow}
\usepackage{booktabs} %

\usepackage{algorithm} 
\usepackage{algpseudocode}

\usepackage[table,xcdraw]{xcolor}
\usepackage{enumitem}

\usepackage{dcolumn}
\newcolumntype{d}[1]{D{.}{.}{#1}}
\usepackage{xpatch}
\xpatchcmd{\proof}{\itshape}{\normalfont\proofnamefont}{}{}
\newcommand{\proofnamefont}{} %
\renewcommand{\proofnamefont}{\bfseries}

\usepackage{bm}
\DeclareMathOperator*{\argmax}{arg\,max}
\DeclareMathOperator*{\argmin}{arg\,min}

\newcommand{\bx}{\mathbf{x}}

\newcommand{\bbeta}{\bm{\beta}}

\newcommand{\bgamma}{\bm{\gamma}}

\newcommand{\eeta}{\bm{\eta}} 
\newcommand{\mR}{\mathbb{R}}

\newcommand{\calI}{\mathcal{I}}
\newcommand{\cF}{\mathcal{F}}
\newcommand{\calD}{\mathcal{D}}
\newcommand{\calQ}{\mathcal{Q}}

\newcommand{\rd}{\mathrm{d}}
\def\wh{\widehat}
\def\wt{\widetilde}

\def\beqr{\begin{eqnarray}}
\def\eeqr{\end{eqnarray}}
\def\beqrs{\begin{eqnarray*}}
\def\eeqrs{\end{eqnarray*}}

\newcommand{\trans}{^{\mbox{\tiny{T}}}}
\newtheorem{theorem}{\bf Theorem}%
\newtheorem{lemma}{\bf Lemma}

\newtheorem{proposition}{Proposition}

\usepackage{cleveref}

\newtheorem{assumption}{Assumption}       
 
\crefname{assumption}{assumption}{assumptions}
\Crefname{assumption}{Assumption}{Assumptions}

\crefname{theo}{Theorem}{Theorems}
\crefname{lemma}{Lemma}{Lemmas}
\crefname{coro}{Corollary}{Corollaries}
\crefname{prop}{Proposition}{Propositions}
\newtheorem{remark}{Remark}

\newcommand{\inner}[2]{\langle #1, #2 \rangle}

\newcommand{\abs}[1]{\vert#1\vert}

\newcommand{\opt}{\text{opt}}

\usepackage[final]{microtype}

\usepackage{authblk}        

\makeatletter
\newcommand{\samethanks}[1][\value{footnote}]{\footnotemark[#1]}
\makeatother

\title{Scalable and Distributed Individualized Treatment Rules for Massive Datasets}

\author[1,2]{Nan Qiao\thanks{These authors contributed equally.}}
\author[3]{Wangcheng Li\samethanks}
\author[1,2]{Jingxiao Zhang\thanks{Corresponding author: \href{mailto:zhjxaioruc@163.com}{zhjxaioruc@163.com}}}
\author[4]{Canyi Chen\thanks{Corresponding author: \href{mailto:canyic@umich.edu}{canyic@umich.edu}}}

\affil[1]{Center for Applied Statistics and School of Statistics, Renmin University of China, Beijing, China}
\affil[2]{School of Statistics, Renmin University of China, Beijing, China}
\affil[3]{School of Statistics, Beijing Normal University, Beijing, China}
\affil[4]{Department of Biostatistics, University of Michigan, Ann Arbor, United States}
\date{}   
\begin{document}

  \maketitle

\begin{abstract} 
Synthesizing information from multiple data sources is crucial for constructing accurate individualized treatment rules (ITRs). However, privacy concerns often present significant barriers to the integrative analysis of such multi-source data. Classical meta-learning, which averages local estimates to derive the final ITR, is frequently suboptimal due to biases in these local estimates. To address these challenges, we propose a convolution-smoothed weighted support vector machine for learning the optimal ITR. The accompanying loss function is both convex and smooth, which allows us to develop an efficient multi-round distributed learning procedure for ITRs. Such distributed learning ensures optimal statistical performance with a fixed number of communication rounds, thereby minimizing coordination costs across data centers while preserving data privacy. Our method avoids pooling subject-level raw data and instead requires only sharing summary statistics. Additionally, we develop an efficient coordinate gradient descent algorithm, which guarantees at least linear convergence for the resulting optimization problem. Extensive simulations and an application to sepsis treatment across multiple intensive care units validate the effectiveness of the proposed method.
\end{abstract}

\noindent%


\section{Introduction}

Constructing precise Individualized Treatment Rules (ITRs) is crucial for analyzing data from multi-center clinical trials. 
ITRs optimize health outcomes by tailoring treatments to individual patient characteristics. For instance, \citet{zhang2012estimating} demonstrated that younger patients with primary operable breast cancer and lower progesterone receptor (PR) levels benefit more from treatment with L-phenylalanine mustard and 5-fluorouracil (PF) compared to the combination of PF and tamoxifen (PFT). 

A variety of batch-learning methods have been developed to estimate the optimal ITR by pooling data from multiple sources. These methods include outcome-weighted classification using support vector machines \citep{zhao2012estimating}, C-learning based on the 0-1 loss function \citep{zhang2012estimating}, and regression-based approaches incorporating Adaboost \citep{kang2014CombiningBiomarkersOptimize}, among others. Few methods incorporate the double robustness property inherent in C-learning \citep{zhang2018c}, which reformulates ITR estimation as a classification problem. C-learning has gained considerable attention due to its ability to produce parsimonious and interpretable ITRs \citep{wu2023transfer}.

Despite these advances, the development of ITRs through batch learning is hindered by significant challenges related to data-sharing restrictions across clinical centers. These challenges, stemming from concerns over data security and privacy as well as the burdensome institutional review board approval processes, are not easily mitigated. For instance, in the Observational Health Data Sciences and Informatics network, local health databases are only permitted to share aggregated information rather than patient-level data with the coordinating center  \citep{bayle2023communication}. Consequently, there is a pressing need for distributed learning approaches that leverage multi-center data while preserving patient-level data privacy.

Distributed learning has emerged as a promising solution to these challenges. For example, a meta-estimate of the ITR is obtained by computing an inverse-variance weighted average of site-specific ITRs derived from individual data sources \citep{cheng2022meta}. These methods utilize site-specific summary statistics, rather than individual-level data, to aggregate ITRs. However, local ITR estimates as a nonlinear function of local data are often biased, and averaging these biased estimates can lead to suboptimal final ITRs \citep{braverman2016communication}.
Multi-round methods are hence developed to mitigate the bias for optimal estimate  \citep{michaeli.jordan2019communication,tan2022Communication}. However, the non-smooth nature of the 0–1/hinge loss functions, coupled with the reuse of data in constructing the weights for C-learning, poses substantial challenges for developing valid distributed method. A comprehensive literature review and comparative analysis are provided in Web Appendix A.

In this paper, we aim to enhance both the estimation accuracy and computational efficiency of interpretable ITR estimation in multi-center studies. %
Our key contributions are as follows:

\begin{itemize}[labelwidth=0pt]
\item {\it A new smooth and convex loss function.} To address the challenge of minimizing the weighted classification error for ITR estimation, we introduce a smoothed and convex surrogate for the non-smooth hinge loss with a convolution technique  \citep{tan2022Communication}.   This helps develop a full-sample convoluted smoothing estimator (FCE). 
\item {\it A multi-round distributed learning procedure.} To accommodate distributed data, we use the approximate Newton method of \citet{michaeli.jordan2019communication} to develop a multi-round distributed algorithm. Each round involves exchanging only summary-level gradient information across data sites. This iterative strategy linearly reduces estimation error and converges to the FCE after a finite number of communication rounds. 
\item {\it A linear convergent optimization algorithm.} We propose a generalized coordinate descent algorithm tailored for the smooth, convex objective function, ensuring linear convergence in optimization. Compared to Newton methods, which require expensive Hessian inversion, our approach significantly reduces computational complexity while maintaining scalability and robust empirical performance.
\item \textit{Non-asymptotic bounds for value loss in both pooled and distributed settings.} The reuse of data for estimating weights introduces troublesome dependence in the objective function. The theoretical analysis is more challenging.  
Using empirical process theory, we derive non-asymptotic upper bounds on the value loss of the estimated ITR. The bound decomposes the total error into five interpretable components corresponding to distinct sources of statistical and algorithmic error. We further extend these results to the distributed setting.

\end{itemize}

In the rest of the article, we first introduce the proposed smoothed estimator using full sample in \Cref{section:FCE} and develop an efficient distributed algorithm in \Cref{section:DCE}.
We then present the theoretical results in \Cref{section:theory}.
The proposed methods are evaluated via a numerical simulation and a real-world application on Sepsis treatment across intensive care units in Sections~\ref{section:simulation} and~\ref{section:real_data}, separately. We finally provide a conclusion in \Cref{section:conclusion}.

\section{Proposed Method}\label{section:FCE}
\subsection{Setup}
Consider $M$ sites have collected a sample $\{(Y_i, A_i, \bx_i ) \}_{i=1}^N$ of size $N$, where $\bx_i = (X_{i1},\ldots, X_{ip}) \in  \mR^{p} $ denotes a vector of individual characteristics, $A_i \in \{0,1\}$ represents the binary treatment, and $Y_i \in \mR$ denotes the (clinical) outcome of interest, with larger values indicating greater benefit. 
The data maybe generated from either randomized trials or observational studies. 
Let $\calI_k$ denote the index set of individuals at $k$-th site, 
with $\calD_k = \{(Y_i, A_i, \bx_i)\}_{i \in \calI_k}$ representing the corresponding local sample. 
It follows that $\cup_{k = 1}^M \calI_k = \{1, \ldots, N\}$. 
For simplicity, we assume that the data are evenly distributed as $\abs{\calI_k} = N/M = n$, for $k = 1, \ldots, M$.

An ITR is an assignment rule based on individual characteristics, which is formulated as a map $d: \mR^p \to \{0,1\}$ from the space of characteristics to the treatment space. 
Let $Y(A)$ denote the potential outcome under treatment $A$. The potential outcome under any ITR $d(\cdot)$ is defined as $Y(d)= Y(1) \I\{d(\bx)= 1\} + Y(0) \I\{d(\bx)= 0\}$, where $\I(\cdot)$ is the indicator function. Furthermore, define $Q(\bx, A) := \mE\{Y(A) \mid \bx\}$ as the conditional expectation of potential outcomes given $\bx$, and $\delta^*(\bx) := Q(\bx, 1) - Q(\bx, 0)$ as the conditional treatment effect.

Throughout the paper, we impose the following identification assumptions, which are standard and well-established in the fields of causal inference and treatment regime \citep{Imbens2015}: 
(i) Consistency: $Y= Y(1) \I(A=1)+Y(0) \I(A=0)$; 
(ii) Positivity: for some constant $c_1, c_2\in(0, 1)$, the propensity score $\pi_A(\bx) :=  P(A=1 \mid \bx)\in[c_1, c_2]$ for all $\bx\in \mR^p$;
(iii) No unmeasured confounders: $A \perp Y(a) \mid X$ for all $a$.

\subsection{Convolution-smoothed optimal ITR estimation}
In this subsection, we will propose the convolution-smoothed optimal ITR estimator when the full sample $\{(Y_i, A_i, \bx_i ) \}_{i=1}^N$ is available.

Denote the marginal mean outcome as $V(d)=\mE\{Y(d)\} = \mE [Q(\bx, 1) \I\{d(\bx)= 1\}+Q(\bx, 0) \I\{d(\bx)= 0\}]$.
Then, the optimal ITR, $d^\opt(\bx)$, is the rule which maximizes $V(d)$ as $d^\opt(\bx) = \argmax_{d}V(d)$.
\cite{zhang2012estimating} suggested learning the optimal ITR by minimizing a weighted misclassification error:
\begin{eqnarray}\label{eq:decision1}
    d^{\text{opt}}(\bx) 
    = 
    \argmin_d  
    \mE  \bigg( 
        |\delta^*(\bx)| 
        \I \big[ 
            \I \{ \delta^*(\bx) > 0 \} \ne d(\bx) 
        \big] 
    \bigg), 
\end{eqnarray}
where $|\delta^*(\bx)|$ is the weight and $ \I \{ \delta^*(\bx) > 0 \} \ne d(\bx)$ represents the misclassification event.

This paper considers ITR with form $d(\bx) = \I\{f(\bx) > 0\}$ and $d^{\text{opt}}(\bx) =  \I\{f^{\text{opt}}(\bx) > 0\}$, where $f (\cdot) \in \cF$ is a decision function and $\cF$ is a user-specified function family.
With the full sample, we can estimate $f^{\text{opt}}(\bx)$ by minimizing the empirical counterpart of \eqref{eq:decision1}:
\begin{eqnarray}\label{eq:decision}
    \hf = 
    \argmin_{f \in \cF} 
    \sum_{i=1}^{N}  
    |\widehat{\delta}_{m,i}| 
    \I \Big[  
        \I (\widehat{\delta}_{m,i} > 0)  \ne \I\{f(\bx_i) > 0\} 
    \Big]  
    = 
    \argmin_{f  \in \cF}
    \sum_{i=1}^{N}
    |\widehat{\delta}_{m,i}| 
    \psi \Big\{ 
        \wh Z_{m,i}
        \times
        f(\bx_{i}) 
    \Big\},
\end{eqnarray}%
where $\widehat{\delta}_{m}(\cdot)$ is an estimator for $\delta^*(\cdot)$, 
$\widehat{\delta}_{m,i} = \widehat{\delta}_{m}(\bx_i)$, 
$\wh Z_{m,i} = 2 \I ( \widehat{\delta}_{m,i} > 0) - 1$ is the pseudo-category, 
$\psi(t) = \I(t \le 0) $ is the 0-1 loss, 
and the subscript $m$ represents the working model.
Then, the estimated optimal ITR is $\wh d^\opt(\bx) = \I \{\hf (\bx) > 0\}$.
To derive an interpretable ITR, we set $\cF = \mathcal{B} = \{f(\bx; \bbeta) = \beta_0+\bbeta_{1}\trans\bx,~\bbeta = (\beta_0,\bbeta_{1}\trans)\trans \in \mathbb{R}^{p+1}\}$.
It can be generalized to handle non-linear decision rules by adopting a kernel function in the context of reproducing kernel Hilbert space \citep{cortes1995SupportvectorNetworks}.

Optimizing problem \eqref{eq:decision} is particularly challenging because the 0-1 loss function in the objective is neither smooth nor convex. 
An appealing alternative is the hinge loss $\varphi(t) = \max(1-t, 0)$, 
which is widely adopted in the field of Support Vector Machine (SVM) problems \citep{zhao2012estimating, liu2018augmented}. Then, the surrogate objective function is
\begin{eqnarray}
  \frac{1}{N} \sum_{i=1}^{N}  |\widehat{\delta}_{m,i}| \times \varphi \{\wh Z_{m,i}   f(\bx_i; \bbeta)\} + \lambda \|\bbeta_{1}\|_2^2 :=  \wh\calQ(\bbeta) +  \lambda \|\bbeta_{1}\|_2^2, 
    \label{eq:decision2}
\end{eqnarray}%
where $\|\bbeta_{1}\|_2$ is the ridge penalty used for avoiding  overfitting, and $\lambda>0$ is a tuning parameter.
However, the hinge loss is non-smooth as well, which poses challenges in developing efficient algorithms and conducting theoretical analysis.
To address the issue,
we suggest applying the convolution-smoothing procedure \citep{fernandes2021smoothing,tan2022Communication,tan2022high} to the hinge loss, where the smoothed loss is $\varphi_h(t) = \int \varphi (u) K_h(u-t) \rd u$,
$K(\cdot)$ is a kernel function, $K_h(t):= K(t/h)/h$, and $h > 0$ is the bandwidth.
Throughout the paper, we use the Epanechnikov kernel $K^E(u)=3/4 (1-u^2)  \I(-1\leq u\leq1)$.
We will also introduce several commonly used kernels and the explicit form of smoothed loss in Web Appendix C.1.

The smoothed hinge loss is now twice differentiable with $\varphi^\prime_h(t) = - \int_{-\infty}^{(1-t)/h} K (u) \rd u$ and $\varphi^{\prime\prime}_h(t) =  K\{(1-t)/h\}/h$.
Replacing the hinge loss with its smoothed version, we derive the smoothed objective function
\begin{eqnarray}\label{eq:estimatefwithqh}
    \wh\calQ_h(\bbeta) = \frac{1}{N} \sum_{i = 1}^N |\widehat{\delta}_{m,i}| \times \varphi_h \{\wh Z_{m,i}   f(\bx_i; \bbeta) \}
    , \quad \mbox{and }
    \wh\bbeta 
    = 
    \argmin_{
        \bbeta \in \mR^{p+1}
    } 
    \wh{\calQ}_h(\bbeta) + \lambda \|\bbeta_{1}\|_2^2.
\end{eqnarray}
Then, the estimated optimal ITR is $\wh d^\opt (\bx_i) = \I \{f(\bx_i;\wh \bbeta) > 0\}$.

\begin{remark}\label{remark:AIPWE}
We suggest using a doubly robust estimator for $\delta^*$ to safeguard potential model misspecification. Specifically, we consider the Augmented Inverse Probability Weighted Estimator (AIPWE) \citep{zhang2012estimating,zhao2015doubly,pan2021improved}. {Let ${\pi}_A(\bx_i;\bgamma)$ and $Q(\bx_i, a;\eeta_a)$ be working models of $\pi_A(\bx)$ and $Q(\bx, a)$ indexed by $\bgamma$ and $\eeta_a$ for $a\in\{0, 1\}$.}  For $i = 1,\ldots,N$, the AIPWE for $\delta^*(\bx_i)$ is given by:
\begin{eqnarray}
\label{eq:delta_est}
\resizebox{0.91\hsize}{!}{%
$\widehat{\delta}_{m,i} =   
    {\left[\frac{\I(A_i=1)\{Y_i-{Q(\bx_i, 1;\wh\eeta_1)}\}}{{\pi}_A(\bx_i;\wh\bgamma)}+ {Q(\bx_i, 1;\wh\eeta_1)}\right] }  - \left[\frac{\I(A_i=0)\{Y_i-{Q(\bx_i, 0;\wh\eeta_0)}\}}{1-{\pi}_A(\bx_i;\wh\bgamma)}+{Q(\bx_i, 0;\wh\eeta_0)}\right].$%
    }
\end{eqnarray}
The double robustness means that $\widehat{\delta}_{m,i}$ is consistent for $\delta^*(\bx_i)$, as long as one of the nuisance working models is correctly specified and consistently estimated  \citep{pan2021improved}. Let $\wt{\bx} = (1, \bx\trans)\trans$. In particular, we can assume a linear regression model ${Q(\bx, a;\eeta_a)}$ for  $Q(\bx, a)$ and a logistic regression model ${\pi}_A(\bx; \bgamma) = \{ 1 + \exp(- \bgamma\trans\wt{\bx})\}^{-1}$ for $\pi_A(\bx)$, and take $\wh\bgamma$ and $\wh\eeta_a$ as the maximum likelihood estimators.

\end{remark}

\section{Optimization}\label{section:DCE}

\subsection{Distributed learning procedure}
In this section, we will propose the Distributed Convolution-smoothed Estimate (DCE) for ITRs.
With the convolution-smoothed objective function introduced in Section~\ref{section:FCE}, we can apply the gradient-based distributed algorithm proposed by \cite{michaeli.jordan2019communication}.

To be specific, let $\wh\calQ_{k, h}(\bbeta) = n^{-1} \sum_{i \in\calI_k} [|\widehat{\delta}_{m,i}| \times \varphi_h \{\wh Z_{m,i}   f(\bx_i; \bbeta) \}]$ be the risk function at $k$-th site. 
At iteration $t$, the first (central) site %
broadcasts $\wt\bbeta^{(t - 1)}$ obtained at iteration $t - 1$ to all local sites. 
For $k = 1, \ldots,M$, the $k$-th local site then computes gradients $\nabla\wh\calQ_{k, h}(\wt\bbeta^{(t - 1)})$, which are then transmitted back to the central site for constructing the global gradient $\nabla \widehat{\calQ}_{h}(\wt\bbeta^{(t - 1)}) = M^{-1} \sum_{k = 1}^{M} \nabla \widehat{\calQ}_{k, h}(\wt\bbeta^{(t - 1)})$. The first site further updates the estimate by solving the following optimization problem on the central site,
\begin{eqnarray}
    \label{optimization:distributed}
    \wt\bbeta^{(t)} = \argmin_{\bbeta \in \mR^{p+1}}
    \wh\calQ_{1, b}(\bbeta) - \inner{\nabla\wh\calQ_{1,b}(\wt\bbeta^{(t - 1)}) - \nabla\wh\calQ_{h}(\wt\bbeta^{(t - 1)})}{\bbeta - \wt\bbeta^{(t - 1)}} + \lambda\|\bbeta_{1}\|_2^2,
\end{eqnarray} %
where $ \wh\calQ_{1,b}(\bbeta) =  n^{-1}\sum_{i \in \calI_1} [ \abs{\widehat{\delta}_{m,i}} \times \varphi_b \{\wh Z_{m,i}  (\beta_0 + \bbeta_{1}\trans \bx_{i})\} ]$, and $b\geq h>0$ are the local and global bandwidths respectively.
We summarize the above procedure in Algorithm \ref{alg:distributed}.

\begin{algorithm}[!ht]
    \caption{{Distributed Optimal ITRs Estimation via Convoluted Smoothing.}} 
  \label{alg:distributed}
    \begin{algorithmic}[1]%
        \Require Data $\{(Y_{i}, A_i, \bx_i)\colon i\in\calI_{k}\}$ for $k=1,\ldots, M$, stored on $M$ local sites, 
        max iterations $T$, bandwidths $b\geq h>0$, initialization $\widetilde{\bbeta}^{(0)}$,
        regularization parameter $\lambda>0$.
    \For{$t = 1, \ldots, T$}
        \State Send $\wt\bbeta^{(t - 1)}$ to all local sites.
        \For{$k = 1, \ldots, M$}
        \State Compute $\nabla \widehat{\calQ}_{k, h}(\wt\bbeta^{(t - 1)})$ on the $k$-th local site, and send it back to the first site. 
        \EndFor 
        \State Compute the global gradient $\nabla \widehat{\calQ}_{h}(\wt\bbeta^{(t - 1)}) = M^{-1} \sum_{k = 1}^{M} \nabla \widehat{\calQ}_{k, h}(\wt\bbeta^{(t - 1)})$, and update $\wt\bbeta^{(t)}$ by minimizing \eqref{optimization:distributed} %
        via \Cref{alg:coordinate_descent}
        on the central site.
        \EndFor%
        \Ensure $\wt\bbeta^{(T)}$.
    \end{algorithmic}
\end{algorithm}

\begin{remark}
    In the distributed scenario, to save communication costs, we recommend using the divide-and-conquer estimators   $\wh\bgamma = \frac{1}{M}\sum_{k = 1}^{M} \wh\bgamma_k$ and $\wh\eeta_a =  \frac{1}{M}\sum_{k = 1}^{M} \wh\eeta_{k,a}$, where $\wh\bgamma_k = \argmax_{\bgamma} \sum_{i\in \calI_k} A_{i} \log\pi_A(\wt{\bx}_{i};\bgamma) + (1 - A_{i}) \log\{{1-\pi_A(\wt{\bx}_{i};\bgamma)}\}$, and $\wh\eeta_{k,a} = \arg\min_{\eeta} \sum_{\{i \in \calI_k ; A_{i} = a\}}(Y_i - \eeta\trans\wt{\bx}_i)^2$. A more cost-effective method is to use $\wh\bgamma_1$ and $\wh\eeta_{1,a}$ calculated only on the central site. 
\end{remark}

\subsection{Generalized coordinate gradient descent on the central site}
Newton method can be used to solve the problem \eqref{optimization:distributed}.
However, it is computationally expensive since it involves inverting the Hessian matrix at each iteration.
To address the issue, we develop the generalized coordinate gradient descent optimization algorithm for solving the problem \eqref{optimization:distributed}, which guarantees a linear convergence and offers a simple implementation. We first show that the first-order derivative of $\varphi_b(\cdot)$ is Lipschitz continuous. 
\begin{lemma}[Lipschitz continuity]
\label{lemma:lipschitz}
	The first-order derivatives of the convolution-smoothed loss, $\varphi_b^\prime(u)$, 
    are Lipschitz continuous as $\abs{\varphi_b^\prime(u_1) - \varphi_b^\prime(u_2)} < c_b \abs{u_1 - u_2}$, 
	with Lipschitz constant $c_b^E = 3/(4b)$ for Epanechnikov kernel. Accordingly, the convoluted hinge loss using Epanechnikov kernel admits the following quadratic majorization, %
	\begin{eqnarray}
 \label{eq:lipschitz}
		\varphi_b(u_1)\leq \varphi_b(u_2) + \varphi_b^\prime(u_2)(u_1 - u_2) + c_b(u_1 - u_2)^2/2,
	\end{eqnarray}
	where $\varphi_b$ is instantiated by $\varphi_b^E$, and $c_b$ is the corresponding Lipschitz constant. %
\end{lemma} %

Based on \Cref{lemma:lipschitz}, we combine the coordinate gradient descent and majorize-minimization techniques to achieve a fast linear convergence.
Without loss of generality, we standardize the observations on the central site such that $\sum_{i \in \calI_1}X_{ij} = 0$ and $\sum_{i \in \calI_1}X_{ij}^2/n = 1$ for $j = 1, \ldots, p$. %
In a coordinate-wise manner, suppose $\beta_0, \beta_1, \ldots, \beta_{j - 1}$ have been updated and we now update $\beta_j$. Let $\wt\bbeta$ be the current solution and $v_i =  \wh Z_{m,i}\bx_i\trans\wt\bbeta$. 
To update $\beta_j$, instead of solving the coordinate-wise update function \eqref{optimization:distributed},  we opt to minimize its majorization function: 
\begin{eqnarray*}
& & \min_{\beta_j \in \mR} \Bigg[ \frac{1}{n} \sum_{i \in\calI_1} |\wh\delta_{m,i}| \varphi_b(v_i)  +
   \frac{1}{n} \sum_{i \in\calI_1} |\wh\delta_{m,i}| \varphi_b^\prime(v_i) \wh Z_{m,i} X_{ij}\big(\beta_j - \wt{\beta}_j\big) + {\color{black} C_{b,j}} \big(\beta_j - \wt\beta_j\big)^2/2
      \\
& &\hspace{2cm} - \big\{\nabla\wh\calQ_{1,b}(\wt\bbeta^{(t-1)}) - \nabla\wh \calQ_{h}(\wt{\bbeta}^{(t-1)}) \big\}_j\big(\beta_j - \wt{\beta}_j\big) + \lambda  \beta_j^2 \Bigg].
\end{eqnarray*}
The minimizer of the above display is 
\begin{eqnarray}\label{update:coordinate_descent}
 \frac{{\color{black} C_{b,j}}  }{2\lambda  +
 {\color{black} C_{b,j}}  }\wt\beta_j - 
 \frac{1}{n(2\lambda   + 
 {\color{black} C_{b,j}}  )}\sum_{i \in \calI_1} |\wh\delta_{m,i}| \varphi_h^\prime\big(v_i\big) \wh Z_{m,i} X_{ij}   +  \frac{1}{2\lambda  + {\color{black} C_{b,j}}  } \big\{\nabla\wh\calQ_{1,b}(\wt\bbeta^{(t-1)}) -
 \nabla\wh\calQ_{h}(\wt\bbeta^{(t-1)})\big\}_j,%
\end{eqnarray}
where ${\color{black} C_{b,j} = c_b\sum_{i \in \calI_1}\abs{\wh\delta_{m,i}}X_{ij}^2/n}$.
Likewise, $\beta_0$ is updated as above with $\lambda = 0$. We summarize the above generalized coordinate descent algorithm in \Cref{alg:coordinate_descent}. Note that our \Cref{alg:coordinate_descent} runs entirely on the central site and hence incurs no extra communication costs. The convexity and smoothness of the new loss function empower a linear convergence to \Cref{alg:coordinate_descent}.

\begin{lemma}\label{lemma:GCD}
    The sequence produced by the generalized coordinate descent \Cref{alg:coordinate_descent} converges at least linearly to an optimal solution of \eqref{optimization:distributed}.
\end{lemma}

\begin{algorithm}[!htbp]
    \caption{Generalized Coordinate Descent Algorithm for Optimization on Central Site.}
  \label{alg:coordinate_descent}
    \begin{algorithmic}[1]%
        \Require Data $\{(Y_{i}, A_i, \bx_{i})\colon i\in\calI_{k}\}$ for $k=1,\ldots,M$,
    the maximum number of iterations $T_{cd}$ for the coordinate descent, the bandwidths $b\geq h>0$,
    and the regularization parameter~$\lambda$.
        \State Set the initial point $\wt\bbeta = \wt\bbeta^{(t - 1)}$.
        \For{$d = 1, \ldots, T_{cd}$}%
        \For{$j = 0,\ldots, p$}%
        \State Update $\wt\beta_j$ as \eqref{update:coordinate_descent}. 
        \EndFor%
        \EndFor%
        \Ensure $\wt\bbeta$
    \end{algorithmic}
\end{algorithm}

\section{Theoretical Results}\label{section:theory}
{\color{black} In this section, we present theoretical guarantees for the proposed algorithms. First, in \Cref{ssec:fullsamplepoolingestimator}, we establish an upper bound on the value loss of the estimated ITR from the full samples relative to the true optimal ITR. 
Then, in \Cref{ssec:distributedestimator}, we demonstrate that the distributed estimator achieving asymptotically equivalent value loss to the full-sample estimator within a logarithmic number of communication rounds,  
}

Recall that $\delta^*(\bx) = Q(\bx,1) - Q(\bx,0)$ represents the true CTE, which is estimated by a working model as in \Cref{remark:AIPWE}.
In the following analysis, we do not require the estimator $\hde_m$ to be consistent to $\delta^*$; instead, it suffices for $\hde_m$ to converge to some $\delta^*_m$ corresponding to the specified working models. Under the principle of double robustness, as long as one of the working models is correctly specified, $\delta^*_m$ will be equivalent to $\delta^*$. 
Additionally, we denote pseudo-category as $Z(\bx) = 2\I\{\delta(\bx) > 0 \}-1$, then $Z^*$, $Z_m^*$ and $\widehat{Z}_m$ are similarly defined.

Necessary notations:
{\color{black} 
For two sequences of real numbers $\{a_n\}_{n \ge 1}$ and $\{b_n\}_{n \ge 1}$, $a_n \lesssim b_n$ indicates that there exists a constant $C > 0$ independent of $n$ such that $\abs{a_n} \le C b_n$; $b_n \gtrsim a_n$ is equivalent to $a_n \lesssim b_n$; and $a_n \asymp b_n $ is equivalent to $a_n \lesssim b_n$ and $b_n \lesssim a_n$. 
For two sets of random variables
$\{X_n\}$ and $\{Y_n\}$, we write $X_n = O_p(Y_n)$ if for any $\epsilon>0$,
there exists a finite $\delta_{\epsilon}>0$ and a finite $n_\epsilon>0$ such that
$\pr(\abs{X_n/Y_n}>\delta_{\epsilon})<\epsilon$ for any $n>n_\epsilon$, and write $X_n = o_p(Y_n)$ if for any $\epsilon, \delta > 0$, there exists a finite $n_{\epsilon, \delta} > 0$ such that  $\pr(\abs{X_n/Y_n}>\delta) < \epsilon$ for any $n > n_{\epsilon, \delta}$.%

}

\subsection{Statistical guarantees for FCE}\label{ssec:fullsamplepoolingestimator}
To bound the value loss, we impose some necessary technique assumptions as follows. 

\begin{assumption}\label{ass:precisionofdelta}
    $\hde_m(\bx)$ point-wisely converges to $\delta_m^*(\bx)$ in probability.
    {\color{black} The estimation error satisfies that 
    $N^{-1} \sum_{i = 1}^{N} \{ \hde_m(\bx_i) - \delta_m^* (\bx_i) \}^2 = O_p (N^{-2\gamma}) $
    for some constant $0 < \gamma \le 1/2$.}
\end{assumption}

\begin{assumption}\label{ass:lowerbounddelta}
    There exists a constant $\eta_u > 0$ such that $|\delta^*_m(\bx)| \le \eta_u$ almost surely for all $\bx$. 
    {\color{black} It follows that $\mathbb{P} \{ N^{-1} \sum_{i = 1}^{N} |\hde_m(\bx_i)| \le \eta_u\} = 1$.}
\end{assumption}

\begin{assumption} \label{assumption:kernel}
$K(\cdot)$ is a symmetric density function, that is, $K(u) = K(-u)$, $K(u) \ge 0$ for all $u \in \mR$, and $\int_{-\infty}^{\infty} K(u) \md u = 1$. Besides, $K(\cdot)$ is bounded with $\kappa_u := \sup_{u \in \mR} |K(u)| < \infty$. We use the notation $\kappa_q = \int_{-\infty}^{\infty} |u|^q K(u) \md u$ for $q \ge 1$ and assume $\kappa_q < \infty$ for $q = 1,2$.
\end{assumption}

\begin{assumption} \label{assumption:variable_bound} The covariates vector 
     $\bx \in \mR^p$ is bounded, that is $\max_{j} \abs{X_j}\le C$  for some positive constant $C$.
     {\color{black} 
     The smallest eigenvalue of the covariance matrix $\Cov (\bx)$ is bounded away from zero.
     }
\end{assumption}

Assumption \ref{ass:precisionofdelta} implies that under the working models, the estimate $\hde_m$ would converge to a fixed function $\delta_m^*$. It generally holds even working models are mis-specified. Moreover, it requires the convergence rate to be $N^{-\gamma}$, where $\gamma$ is a constant determined by the specified model. For example, $\gamma = 1/2$ typically holds for a parametric model. Assumption \ref{ass:lowerbounddelta} imposes an upper bound on the conditional treatment effect and its estimate, ensuring that no single individual has a dominant impact on the final outcome. Assumption \ref{assumption:kernel} is standard for kernel functions in convolution smoothing literature \citep{tan2022Communication,tan2022high}, encompassing a wide range of common kernel functions, like the uniform, Gaussian, and Epanechnikov kernel functions. Assumption \ref{assumption:variable_bound} is a standard requirement for covariates in related fields.

Recall that $f(\bx)$ is the decision function, and $d(\bx)=\I\{f(\bx) > 0\}$ is the corresponding decision rule. Then we define the pseudo value function $V$ given the conditional treatment effect $\delta$ and the decision function $f$ as:
\begin{eqnarray*}
     V \left(f, \delta \right) 
     = 
     \mE \Big(  
         |\delta| 
         \I \big[
            \I ( \delta > 0 )
            = 
            \I\{f(\bx) > 0\}%
         \big]
    \Big),
\end{eqnarray*}
and $f^* = \argmax_{f \in \cF} V(f,\delta^*)$, where $\cF$ contains all measurable functions from $\mR^{p+1}$ into $\mR$.

In the following content, we aim to establish an upper bound for the value gap, which is defined as $V(f^*, \delta^*) - V(\hf, \delta^*)$. Here $\wh f(\bx) = \wh \beta_0 + \bx\trans\wh\bbeta_1$ is our ITR learned on the full sample. To this end, we introduce a key lemma that can bound the value loss via the excess risk. 
\begin{lemma}\label{lem:equivalentloss}
Let $\calQ(f,\delta) = \mE [ |{\delta}|  \varphi \{ Z  f(\bx; \bbeta)  \} ]$ and $\calQ_h(f,\delta) = \mE [ |{\delta}|  \varphi_h \{Z  f(\bx; \bbeta) \}]$. 
If $f^*_m$ minimizes $\calQ(f,\delta_m)$ over $\cF$ with any model $\delta_m$, including the true model for $\delta^*$, then 
    \begin{enumerate}
        \item $V(f^*_m, \delta_m) = \max_{f \in \cF} V(f, \delta_m)$, that is $f^*_m$ yields the maximum value of $V$. 
        \item For any $f \in \cF$, 
        $
            V(f^*_m,\delta_m) - V(f,\delta_m) \le \calQ(f, \delta_m) - \calQ(f^*_m, \delta_m), 
        $
        that is the value loss is bounded by the exceed risk when using hinge loss. 
    \end{enumerate}
\end{lemma}

We omit the detailed proofs of \Cref{lem:equivalentloss}, which are the same as Theorems 3.1 and 3.2 in \cite{zhao2012estimating}. It states that for any given $\delta_m$, the decision function $f^*_m$, obtained by minimizing the hinge loss $\calQ(f,\delta_m)$ over $\cF$, also maximizes the value function $V(f, \delta_m)$. Meanwhile, the loss in value is bounded by the excess risk associated with the hinge loss. 

Furthermore, we introduce the approximation error here. Define the pseudo solution as 
$\tf = 
\arg \inf_{f \in \mathcal{B}}
\left\{\calQ(f,\delta^*_m) + \lambda \|f\|^2
\right\}$,
where $\|f\|^2 = \|\bbeta_1\|^2_2$ for $f(\bx; \bbeta)\in\mathcal{B}$. Then the approximation error is defined as 
$ a(\lambda) 
=
\inf _{f \in \cF} 
\{
    \calQ_h(\tf, \delta) + \lambda\|\tf\|^2
    -
    \calQ_h(f, \delta)
\}$,
which arises from two sources: replacing the function family $\cF$ with a smaller one $\mathcal{B}$, and the penalty $\lambda \|f\|^2$. Now, we establish the following theorem, decomposing the value gap to several components.

\begin{theorem}\label{the:valuelossbound}
    Suppose that Assumptions \ref{ass:precisionofdelta}-\ref{assumption:variable_bound} hold, 
    $h ,\lambda \rightarrow 0$ and $N \lambda  \rightarrow \infty$ as $N \rightarrow \infty$.
    {\color{black} Then for any $\varrho > 0$, there exists a constant $M_\varrho>0$ such that}
    \begin{equation*}
        \begin{aligned}
            V(f^*,\delta^*) - V(\wh{f}, \delta^*) 
            \lesssim %
            2 \sup \limits_{f \in \cF} 
            \left| 
                V(f, \delta^*) 
                - 
                V(f, \delta^*_m) 
            \right|  
            + h
            + a(\lambda) 
            + 
            {\color{black} \frac{\log N}{N \lambda}}
            + 
            {\color{black} M_\varrho N^{-\gamma} \cdot \lambda^{-1/2}}, 
        \end{aligned}
    \end{equation*}
    {\color{black} with probability at least $1-N^{-1}-\varrho$}, where $\hf$ is the estimate derived in \eqref{eq:estimatefwithqh}.
\end{theorem}

This result decomposes the value gap into five parts. The first term reflects the impact of working model mis-specification. Since we have used the doubly robust AIPWE \eqref{eq:delta_est} to estimate $\delta^*$, this term equals zero as long as one of the working models is correctly specified. The second term is introduced by the smoothing operation, but by choosing $h$ appropriately, we can make it negligible while speeding up computation. The third term, $a(\lambda)$, represents the approximation error.
The fourth term reflects estimation error, related to the complexity of the function family $\mathcal{B}$ when a penalty is applied. The final term arises from substituting the estimated condition treatment effect $\wh{\delta}_m$ for the pseudo true effect $\delta_m^*$.

\subsection{Statistical guarantees for DCE}\label{ssec:distributedestimator}

In what follows, we establish a theoretical guarantee for the distributed estimator in \eqref{optimization:distributed}.
To do so, we first investigate how the distributed error (the error between the pooling estimator, FCE, and distributed estimator, DCE) influences the value loss.
We formalize it as follows:
\begin{proposition}\label{prop:valuelossdistributedcase}
    Let $\wt{f}^{(t)}$ be the estimator in $t$-th iteration. 
    Under same assumptions in Theorem \ref{the:valuelossbound}, 
    it follows that {\color{black}for any $\varrho > 0$, there exists a constant $M_\varrho>0$ such that}
    \begin{equation}
        \begin{aligned}
            V(f^*, \delta^*) - V(\wt f^{(t)},\delta^*)
            \lesssim & 
           ~ \eta_u \|\wt{f}^{(t)} - \hf \| 
            + 
            2 \sup \limits_{f \in \cF} 
            \left| 
                V(f, \delta^*) 
                - 
                V(f, \delta^*_m) 
            \right|  
            + h
            + a(\lambda) 
            \\ & 
            + 
            \frac{p \log N}{N \lambda}
            + 
            {\color{black} M_\varrho N^{-\gamma} \cdot \lambda^{-1/2}}, 
        \end{aligned}
        \label{eq:r_bound_dist}
    \end{equation}
    {\color{black} with probability no less than $1-N^{-1}-\varrho$.}
\end{proposition}
Compared with Theorem \ref{the:valuelossbound}, we see that the additional value loss in {\color{black}the} distributed scenario is bounded by the norm $\|\wt{f}^{(t)} - \wh{f}\|$, which is equivalent to $\|\wt{\bbeta}^{(t)} - \wh{\bbeta}\|_2$ when the linear space is used. Thus, it remains to bound the $L_2$ norm of the linear estimator. 
To this end, we first impose a slightly stronger condition on the estimation error of the treatment effect, which is defined over the sample units on the first site. 
\begin{assumption}\label{ass:precisionofdeltalocally}
    {\color{black} The estimation error satisfies that 
    $n^{-1} \sum_{i \in \calI_1} \{ 
        \hde_m(\bx_i) - \delta_m^*(\bx_i) 
    \}^2 = O_p (N^{-2\gamma})$ for some constant $0 < \gamma \le 1/2$.}
\end{assumption}

Then we establish the theoretical result for the distributed estimator $\wt{\bbeta}^{(t)}$ as follows. 

\begin{theorem}\label{the:constrainupdate}
    Assume that $\max(N^{-\gamma}, n^{-1/2}) \lesssim b \lesssim 1$, $N^{-1/2} \lesssim h$, $\lambda = o(1)$, and Assumptions \ref{ass:precisionofdelta}-\ref{ass:precisionofdeltalocally} hold, {\color{black} then for any $u, \varrho > 0$, there exists a constant $\overline{M}_\varrho > 0$
    such that}
    \begin{equation*}
        \| \wt{\bbeta}^{(t)}  - \wh{\bbeta} \|_2 
        \lesssim
        \bigg(
            \underbrace{
            \sqrt{
                \frac{u}{n b} 
            }
            +
            \sqrt{
                \frac{u}{N h} 
            }
            + 
            b}_{\mathrm{contraction}}
        \bigg)
        \| \wt{\bbeta}^{(t-1)} - \wh{\bbeta} \|_2 
        + 
        \underbrace{\overline{M}_\varrho N^{-\gamma}}_{\mathrm{inherence}}
        , 
    \end{equation*}
    with probability no less than {\color{black} $1-3 e^{-u} - \varrho - n^{-1}$.}
\end{theorem}

Theorem \ref{the:constrainupdate} indicates that the inherent loss induced by substituting $\delta_m^*$ with $\hde_m$ will influence the convergence of distributed algorithm as well.
Aside from it, the proposed distributed algorithm achieves a linear convergence rate. Consequently, after a logarithmic number of communication, the distributed estimator $\wt{\bbeta}^{(t)}$ reaches the same statistical efficiency as the full-sample estimator $\wh{\bbeta}$. Combined with Proposition \ref{prop:valuelossdistributedcase}, we have that the value loss is asymptotically optimal as well. This confirms the efficiency of the proposed method. 
In addition, we suggest taking $h = O(N^{-\gamma} \lambda^{-1/2})$ and $b \asymp n^{-1/3}$ to balance the smoothing bias and computational burden. Please refer to Web Appendix C.2 for details.

\section{Numerical Study}\label{section:simulation}

We perform extensive simulations to evaluate the performance of full-sample and distributed convolution-smoothed optimal ITR estimators proposed in \Cref{section:FCE,section:DCE}. We denote them by FCE and DCE, respectively. We also compare them with the following estimates. 

{\it Evaluated methods.} (i) Avg C-learning: the averaging-based estimator based on local C-learning \citep{zhang2012estimating} estimators with the smooth hinge loss, where the decision functions and propensity scores are estimated independently at local sites; (ii) Full C-learning: global C-learning estimator using all of the available $N = M n$ observations; (iii) Initial: the initial estimator of our DCE using the observation from the central site only. 

{\it Criteria.} Methods are evaluated according to two criteria: (1) Correct classification rate: the percentage of instances where the estimated optimal treatment regime aligns with the theoretically optimal treatment regime. (2) Empirical value function: $\wh V(d) =  \mE_N[Y\cdot\I\{A=d(\bx)\}/\pi_A(\bx)]/ \mE_N[\I\{A=d(\bx)\}/\pi_A(\bx)]$, where $\mE_N$ denotes the empirical average.
A higher correct classification rate and a higher empirical value indicate superior performance. For all simulations, performance based on the two criteria is assessed using a large independent dataset containing 10,000 observations. Each simulation is repeated 100 times. The reported values are averaged over the replicated datasets, with standard deviations provided in parentheses.

{\it Data Generation.} 
We set the full sample size $N = (1000, 3000, 5000)$ and the local sample size $n = (200, 500)$. We generate three independent covariates $\bx_i = (X_{i1}, X_{i2}, X_{i3},X_{i4},X_{i5})$, each from the uniform distribution $U(-1,1)$. 
For the treatment assignment, we consider two ways: (I) Randomized Controlled Trial (RCT): $A$ is independent of $\bx$, with a known randomization probability of $0.5$ as $\mathbb{P}(A = 1) = 0.5$; 
(II) Observational study: the true treatment propensity model is specified as $\mathbb{P}(A = 1 | \bx) = 1/[1+\exp\{-(0.1 + 0.25X_{1} +0.25X_{2})\}]$. %
The clinical outcome generate from $ Y_i|(\bx_i, A_i) = 1 + 2X_{i1} + 3X_{i2}+ 4X_{i3} + 5X_{i4}  + 6X_{i5} + A_i \delta^\ast(\bx_i) + N(0, 0.5^2)$. 

To evaluate the performance of our methods, we consider two categories of data-generating models. One is the scenarios where the true optimal policies are not in linear forms, that is, the scenarios (b) and (d), while the other one considers the true optimal policy is in a linear form as the scenarios (a) and (c), where ${\rm (a)}~\delta^\ast(\bx_i) = X_{i1} + 2X_{i2}+ 3X_{i3} + 4X_{i4} + 5X_{i5}$, ${\rm (b)} ~ \delta^\ast(\bx_i) = 0.4\times |X_{i3}|(1 - X_{i1} - X_{i2})$, ${\rm (c)} ~\delta^\ast(\bx_i) = 1.8\times(0.9 - X_{i1})$, and ${\rm (d)}~ \delta^\ast(\bx_i) = \arctan\{ \exp(1+X_{i1}) - 3X_{i2} - 5\}$.
Note that the second scenario corresponds to a situation where $X_3$ interacts with the treatment but does not affect the treatment selection strategy. The third scenario represents a case where prescriptive variables have weak predictive power for the outcome but are crucial for decision-making.

Tables \ref{tab:aipw} and \ref{tab:aipw_value} show the correct classification rates (CCRs) and mean empirical value for all four scenarios in the observational study. Overall, the results indicate that FCE and DCE exhibit higher CCRs and values compared to the other three methods, particularly when the local sample size is small. Despite the initial estimate of DCE relying solely on the local sample and performing poorly, our DCE still achieves performance comparable to FCE and Full C-learning using full sample and outperforms the other distributed methods like Avg C-learning. Specifically, for the first linear scenario (a), FCE and DCE have comparable performance with Full C-learning. For Scenario (b), our methods outperform Full C-learning and Avg C-learning. This is because our convolution-smoothed techniques are more robust when prescriptive covariates interact qualitatively within the decision function, leading to better classification and, consequently, higher empirical value. For Scenario (c), our methods also surpass Full C-learning and Avg C-learning. As mentioned in the Introduction, relying on an average estimate can yield a suboptimal ITR estimate, particularly when the prescriptive variable has weak predictive power for the outcome but remains critical for decision-making. In Scenario (d), although the true decision function is nonlinear, the flexible modeling of decisions by FCE and DCE provides an advantage over other C-learning methods. Tables D.1 and D.2 in Web Appendix D show the results of the RCT study.  Additional simulation studies with higher-dimensional settings ($p\in\{5, 10, 15, 20\}$) are provided in Tables D.3 and D.4 of Web Appendix D. They are consistent with the observational study findings, confirming the robustness of the method effects.

\begin{table}%
    \centering
    \caption{Correct classification rate along with standard deviations in parentheses of DCE, FCE, Avg C-learning, Initial, and Full C-learning under varying full sample size $N$ and local sample size $n$ when the true treatment propensity model is specified as $\mathbb{P}(A = 1 | X) = 1/\{1+\exp(-(0.1 + 0.25X_{i1} +0.25X_{i2})\}$.}
    \resizebox{\columnwidth}{!}{
   \begin{tabular}{cccccccc}
    \toprule
          & $N$ & $n$ & DCE   & FCE   & Avg C-learning & Initial & Full C-learning \\
    \midrule
    \multirow{6}[0]{*}{Scenario ${\rm a}$} & \multirow{2}[0]{*}{$N=1000$} & $n=200$ & 0.985(0.006) & \textbf{0.986(0.004)} & 0.981(0.008) & 0.958(0.023) & 0.981(0.013) \\
     &     & $n=500$ &\textbf{0.990(0.004)} & 0.987(0.004) & 0.982(0.009) & 0.976(0.011) & 0.982(0.010) \\
          & \multirow{2}[0]{*}{$N=3000$} & $n=200$ & 0.988(0.007) &\textbf{0.992(0.002)} & 0.988(0.005) & 0.962(0.027) & 0.989(0.005) \\
      &    & $n=500$ &\textbf{0.993(0.003)} & 0.992(0.003) & 0.987(0.008) & 0.974(0.012) & 0.986(0.010) \\
          & \multirow{2}[0]{*}{$N=5000$} & $n=200$ & 0.989(0.007) &\textbf{0.994(0.002)} & 0.990(0.004) & 0.958(0.025) & 0.989(0.014) \\
      &    & $n=500$ &\textbf{0.994(0.002)} & 0.994(0.002) & 0.991(0.005) & 0.975(0.016) & 0.990(0.008) \\
    \midrule
    \multirow{6}[0]{*}{Scenario ${\rm b}$} & \multirow{2}[0]{*}{$N=1000$} & $n=200$ & 0.923(0.046) &\textbf{0.940(0.022)} & 0.790(0.089) & 0.784(0.099) & 0.806(0.131) \\
   &       & $n=500$ & 0.942(0.022) &\textbf{0.942(0.022)} & 0.796(0.110) & 0.815(0.122) & 0.807(0.130) \\
     & \multirow{2}[0]{*}{$N=3000$} & $n=200$ & 0.937(0.034) &\textbf{0.965(0.012)} & 0.785(0.077) & 0.798(0.100) & 0.813(0.136) \\
     &     & $n=500$ & 0.965(0.014) &\textbf{0.966(0.013)} & 0.811(0.116) & 0.819(0.120) & 0.824(0.137) \\
      & \multirow{2}[0]{*}{$N=5000$} & $n=200$ & 0.946(0.040) &\textbf{0.971(0.011)} & 0.783(0.072) & 0.792(0.100) & 0.817(0.137) \\
   &       & $n=500$ & 0.970(0.013) &\textbf{0.972(0.011)} & 0.790(0.103) & 0.806(0.122) & 0.822(0.139) \\
    \midrule
    \multirow{6}[0]{*}{Scenario ${\rm c}$} & \multirow{2}[0]{*}{$N=1000$} & $n=200$ & 0.971(0.014) &\textbf{0.982(0.006)} & 0.948(0.034) & 0.913(0.067) & 0.957(0.049) \\
    &      & $n=500$ & 0.981(0.007) &\textbf{0.982(0.006)} & 0.947(0.045) & 0.930(0.067) & 0.945(0.061) \\
          & \multirow{2}[0]{*}{$N=3000$} & $n=200$ & 0.973(0.016) &\textbf{0.989(0.004)} & 0.951(0.023) & 0.912(0.065) & 0.971(0.037) \\
      &    & $n=500$ & 0.987(0.006) &\textbf{0.989(0.004)} & 0.953(0.032) & 0.929(0.067) & 0.965(0.050) \\
          & \multirow{2}[0]{*}{$N=5000$} & $n=200$ & 0.974(0.016) &\textbf{0.992(0.003)} & 0.953(0.018) & 0.915(0.065) & 0.981(0.024) \\
     &     & $n=500$ & 0.988(0.006) &\textbf{0.991(0.003)} & 0.960(0.021) & 0.931(0.065) & 0.979(0.025) \\
    \midrule
    \multirow{6}[0]{*}{Scenario ${\rm d}$} & \multirow{2}[0]{*}{$N=1000$} & $n=200$ & 0.938(0.010) &\textbf{0.942(0.006)} & 0.897(0.082) & 0.808(0.175) & 0.823(0.182) \\
      &    & $n=500$ &\textbf{0.942(0.005)} &\textbf{0.942(0.005)} & 0.872(0.124) & 0.818(0.183) & 0.824(0.182) \\
          & \multirow{2}[0]{*}{$N=3000$} & $n=200$ & 0.940(0.008) &\textbf{0.946(0.004)} & 0.926(0.043) & 0.815(0.174) & 0.828(0.182) \\
    &      & $n=500$ & 0.945(0.004) &\textbf{0.945(0.004)} & 0.895(0.102) & 0.791(0.196) & 0.807(0.192) \\
          & \multirow{2}[0]{*}{$N=5000$} & $n=200$ & 0.940(0.010) &\textbf{0.947(0.003)} & 0.935(0.030) & 0.793(0.183) & 0.831(0.182) \\
    &      & $n=500$ & 0.946(0.003) &\textbf{0.947(0.003)} & 0.889(0.108) & 0.803(0.179) & 0.807(0.193) \\
    \bottomrule
    \end{tabular}
    }
  \label{tab:aipw}%
\end{table}%

\begin{table}%
    \centering
    \caption{Mean empirical value along with standard deviations in parentheses of DCE, FCE, Avg C-learning, Initial, and Full C-learning under varying full sample size $N$ and local sample size $n$ when the true treatment propensity model is specified as $\mathbb{P}(A = 1 | X) = 1/\{1+\exp(-(0.1 + 0.25X_{i1} +0.25X_{i2})\}$.}
    \resizebox{\columnwidth}{!}{
    \begin{tabular}{cccccccc}
    \toprule
          & $N$ & $n$ & DCE   & FCE   & Avg C-learning & Initial & Full C-learning \\
    \midrule
   \multirow{6}[0]{*}{Scenario ${\rm a}$} & \multirow{2}[0]{*}{$N=1000$} & $n=200$ & \textbf{3.958(0.031)} & 3.958(0.032) & 3.955(0.032) & 3.932(0.051) & 3.954(0.033) \\
     &  & $n=500$ & \textbf{3.956(0.035)} & 3.955(0.035) & 3.952(0.036) & 3.948(0.035) & 3.952(0.036) \\
     & \multirow{2}[0]{*}{$N=3000$} & $n=200$ & 3.949(0.038) & \textbf{3.951(0.038)} & 3.950(0.038) & 3.924(0.068) & 3.950(0.037) \\
     &  & $n=500$ & \textbf{3.960(0.041)} & \textbf{3.960(0.041)} & 3.957(0.042) & 3.949(0.042) & 3.957(0.042) \\
     & \multirow{2}[0]{*}{$N=5000$} & $n=200$ & 3.952(0.036) & \textbf{3.954(0.036)} & 3.953(0.036) & 3.923(0.057) & 3.950(0.042) \\
     &  & $n=500$ & \textbf{3.960(0.034)} & \textbf{3.960(0.034)} & 3.959(0.034) & 3.949(0.036) & 3.958(0.034) \\
    \midrule
    \multirow{6}[0]{*}{Scenario ${\rm b}$} & \multirow{2}[0]{*}{$N=1000$} & $n=200$ & \textbf{1.436(0.028)} & {1.442(0.022)} & 1.384(0.053) & 1.372(0.057) & 1.375(0.065) \\
     &  & $n=500$ & \textbf{1.437(0.023)} & \textbf{1.437(0.023)} & 1.375(0.058) & 1.378(0.066) & 1.371(0.065) \\
     & \multirow{2}[0]{*}{$N=3000$} & $n=200$ & 1.436(0.026) & \textbf{1.442(0.025)} & 1.381(0.049) & 1.376(0.058) & 1.373(0.067) \\
     &  & $n=500$ & \textbf{1.441(0.026)} & \textbf{1.441(0.026)} & 1.381(0.058) & 1.373(0.063) & 1.375(0.065) \\
     & \multirow{2}[0]{*}{$N=5000$} & $n=200$ & 1.437(0.027) & \textbf{1.442(0.025)} & 1.381(0.051) & 1.372(0.058) & 1.375(0.069) \\
     &  & $n=500$ & \textbf{1.441(0.022)} & \textbf{1.441(0.022)} & 1.374(0.058) & 1.373(0.063) & 1.375(0.065) \\
    \midrule
    \multirow{6}[0]{*}{Scenario ${\rm c}$} & \multirow{2}[0]{*}{$N=1000$} & $n=200$ & 2.797(0.022) & \textbf{2.801(0.021)} & 2.786(0.033) & 2.740(0.081) & 2.781(0.061) \\
     &  & $n=500$ & \textbf{2.803(0.024)} & \textbf{2.803(0.024)} & 2.783(0.046) & 2.756(0.079) & 2.770(0.069) \\
     & \multirow{2}[0]{*}{$N=3000$} & $n=200$ & 2.796(0.027) & \textbf{2.800(0.027)} & 2.790(0.028) & 2.738(0.085) & 2.790(0.042) \\
     &  & $n=500$ & \textbf{2.796(0.027)} & \textbf{2.796(0.027)} & 2.784(0.034) & 2.747(0.080) & 2.778(0.057) \\
     & \multirow{2}[0]{*}{$N=5000$} & $n=200$ & 2.794(0.027) & \textbf{2.799(0.026)} & 2.790(0.026) & 2.739(0.083) & 2.795(0.028) \\
     &  & $n=500$ & 2.798(0.024) & \textbf{2.799(0.024)} & 2.791(0.025) & 2.753(0.075) & 2.795(0.029) \\
    \midrule
    \multirow{6}[0]{*}{Scenario ${\rm d}$} & \multirow{2}[0]{*}{$N=1000$} & $n=200$ & 1.551(0.024) & \textbf{1.556(0.022)} & 1.499(0.111) & 1.376(0.246) & 1.393(0.249) \\
     &  & $n=500$ & \textbf{1.553(0.025)} & \textbf{1.553(0.025)} & 1.462(0.172) & 1.383(0.253) & 1.391(0.254) \\
     & \multirow{2}[0]{*}{$N=3000$} & $n=200$ & 1.548(0.028) & \textbf{1.555(0.027)} & 1.529(0.059) & 1.379(0.245) & 1.394(0.255) \\
     &  & $n=500$ & \textbf{1.558(0.025)} & \textbf{1.558(0.025)} & 1.493(0.137) & 1.347(0.271) & 1.369(0.264) \\
     & \multirow{2}[0]{*}{$N=5000$} & $n=200$ & 1.552(0.027) & \textbf{1.558(0.024)} & 1.543(0.042) & 1.352(0.253) & 1.400(0.254) \\
     &  & $n=500$ & 1.561(0.020) & \textbf{1.562(0.020)} & 1.487(0.143) & 1.369(0.247) & 1.370(0.267) \\
    \bottomrule
    \end{tabular}
    }
  \label{tab:aipw_value}%
\end{table}%

\section{Sepsis Data Analysis}\label{section:real_data}
In this section, we apply the proposed FCE and DCE on the dataset of sepsis patients from MIMIC-IV sepsis cohort \citep{johnson2023mimic}. %
MIMIC-IV is a freely available database containing deidentified medical records for over 40,000 patients who were admitted to the critical care units at Beth Israel Deaconess Medical Center between 2008 and 2019. Please refer to \cite{johnson2023mimic} for a detailed description.

Sepsis is a life-threatening condition, affecting millions globally each year and resulting in a mortality rate of one in three to one in six among those afflicted \citep{rhee2017incidence}. 
While survival is an important long-term goal, it is not a short-term parameter that physicians can monitor to adjust treatment strategies during hospitalization. Thus we use an alternative metric based on serum lactate levels, an accepted marker for systemic tissue hypoperfusion that can reflect cellular dysfunction in sepsis and be used to predict sepsis mortality \citep{liu2017early}. For adults with sepsis or septic shock, \cite{evans2021surviving} recommends guiding resuscitation by decreasing serum lactate levels in patients with high lactate. Therefore, we use the difference in serum lactate levels before and after ICU admission as the reward ($Y$). A larger decrease in serum lactate is interpreted as a more favorable treatment response. The treatment variable $A$ is defined based on the median dosage of intravenous fluids: patients receiving dosages above the median are assigned to the treatment group ($A = 1$), while those receiving below-median dosages constitute the control group ($A = 0$) \citep{zhou2024federated}.
Patient-level covariates include measurements of weight, temperature, systolic blood pressure, hemoglobin, and potassium level. All covariates are centered and standardized as input. 
Our goal is to estimate a linear ITR as a function of patient-specific covariates for maximizing the reduction in serum lactate levels.

\begin{figure}
    \centering
    \includegraphics[width=\linewidth]{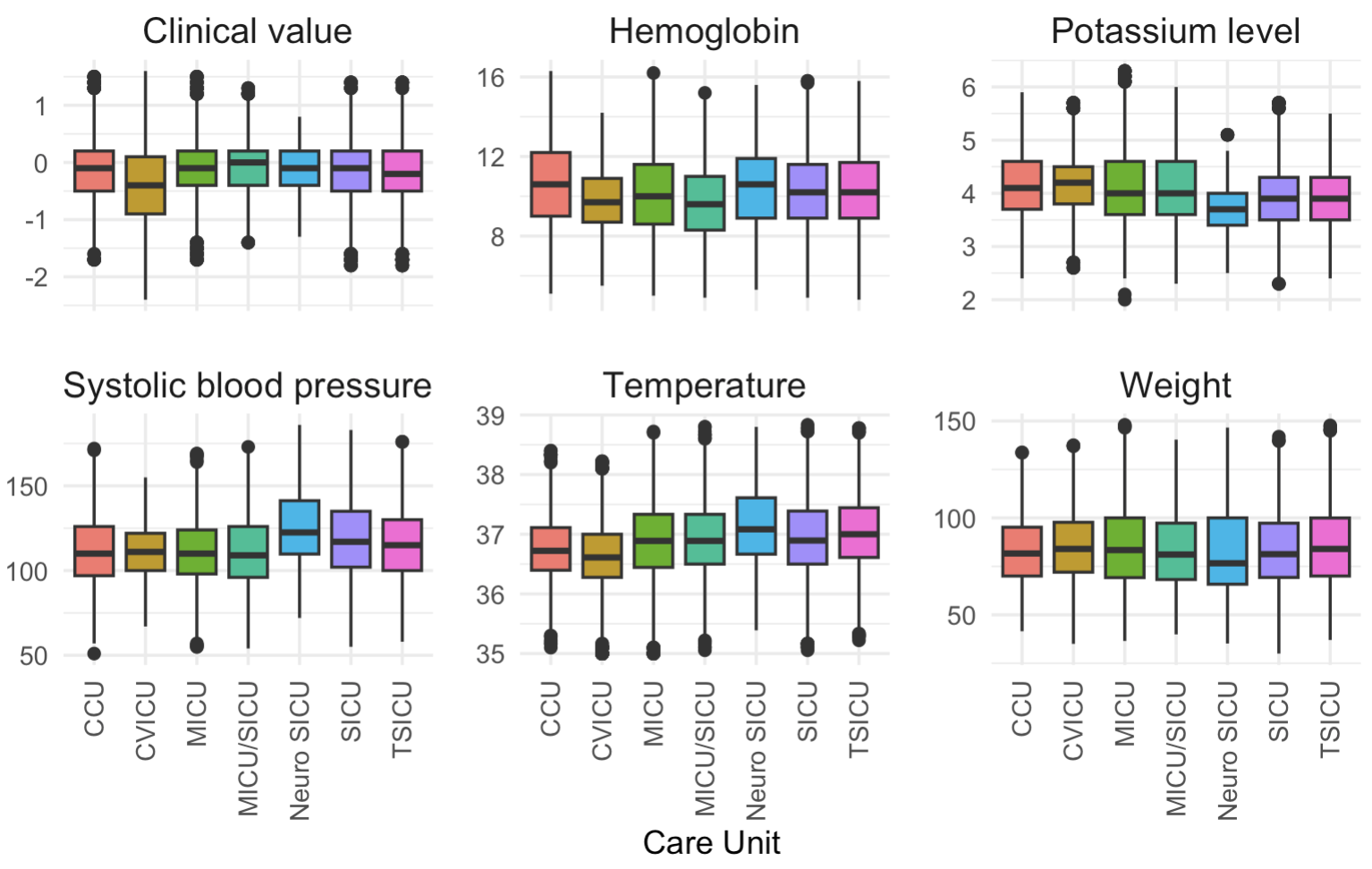}
    \caption{Clinical value (Difference in serum lactate levels before and after ICU admission) and five covariates in seven centers}
    \label{fig:value}
\end{figure}

\begin{figure}
    \centering
    \includegraphics[width = \linewidth]{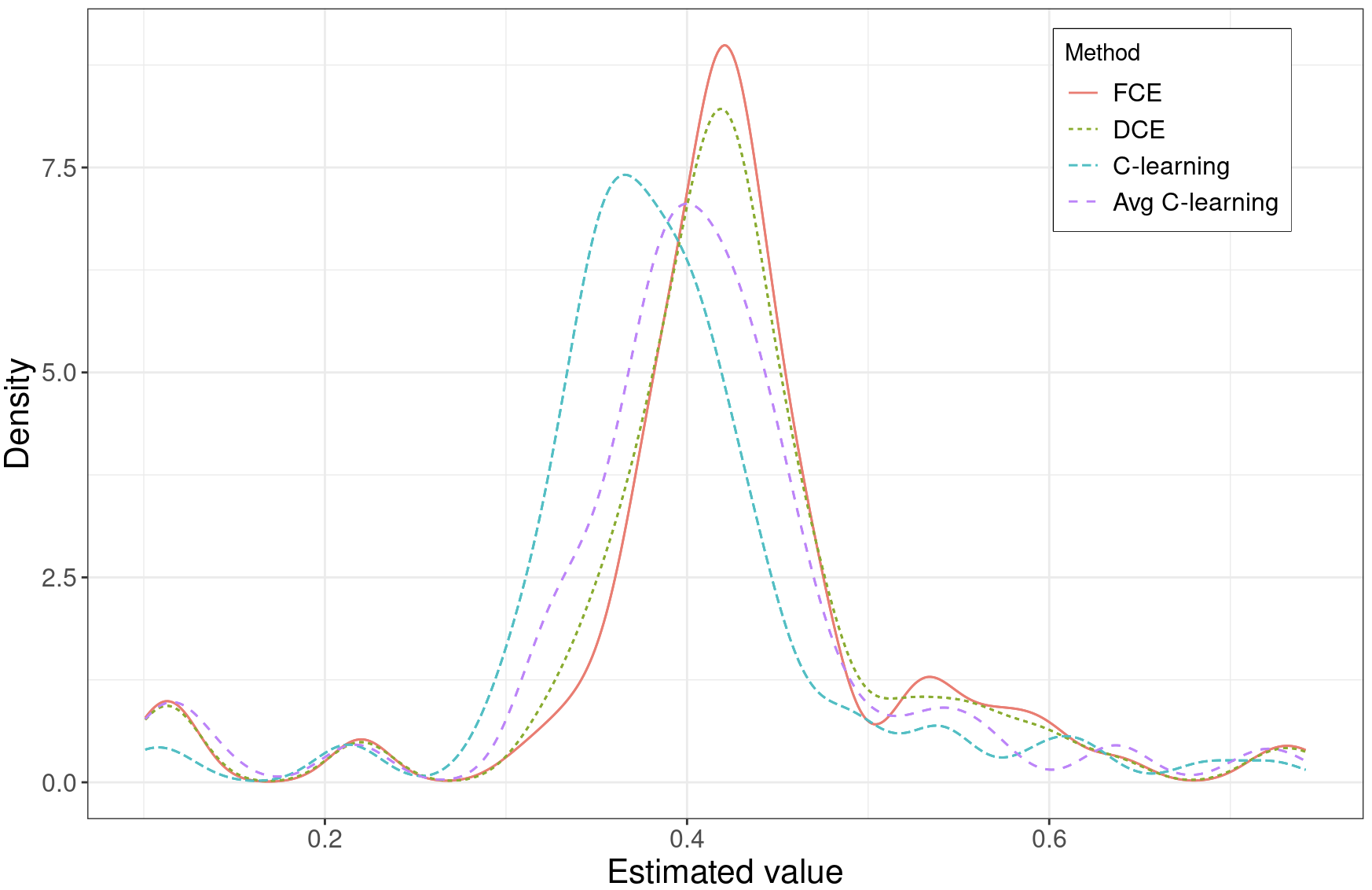}
    \caption{Estimation of empirical value.}
    \label{fig:main}
\end{figure}

{\color{black} Using publicly available SQL queries \citep{JiaTeamSepsisSQL}}%
, we extract data from 7,987 patients admitted to the ICU for the first time across seven different care units.
We ensure that the complete feature variable information was included in our analysis. As our method implies data homogeneity across different centers, Figure \ref{fig:value} illustrates this point.  %
The data is randomly partitioned into training and test sets at a ratio $7:3$.
We repeat the split and prediction process 100 times. Based on this, we plot the density function of empirical value. 
Figure \ref{fig:main} shows that the FCE and DCE methods have sharply defined peaks and higher density values around the estimated value of 0.4, indicating more precise and reliable empirical estimates. In contrast, the C-learning and Avg C-learning methods exhibit broader, less concentrated peaks, with Avg C-learning showing more significant variability and less certainty. 
Therefore, the FCE and DCE methods demonstrate superior performance, making them preferable in distributed settings.

\section{Conclusion} \label{section:conclusion}

In this study, we proposed a novel approach for estimating the optimal ITR in multi-center studies with distributed data. Our method addresses the significant challenges of data privacy, communication efficiency, and computational burden. By reformulating the problem into a classification task and using a convex convolution-smoothed surrogate loss function, we developed a communication-efficient distributed implementation that only requires gradient information exchange. Additionally, we introduced an efficient generalized coordinate descent algorithm to handle the computational complexity typically associated with optimization algorithms. Through simulations and an application to sepsis treatment across intensive care units, we demonstrated the effectiveness and robustness of our approach. Our method provides a practical, privacy-preserving solution for multi-center studies, paving the way for more accurate and effective treatment strategies in precision medicine. In addition, the present study assumes homogeneity in treatment assignment across data sources, which constitutes a notable limitation. Future research may address this by incorporating subgroup analyses, personalized federated learning frameworks, or hierarchical modeling approaches to better accommodate inter-source heterogeneity.


%
%
%
{\color{black}
\section*{ Supplementary Materials}
Web Appendices, technical proofs and additional simulation results referenced in Sections \Cref{section:theory} and \Cref{section:simulation} and code are available with this paper at the Biometrics website on Oxford Academic.
\section*{Data Availability}
The real-world data, MIMIC-IV sepsis cohort, can be requested at \cite{johnson2023mimic}.
}

\vspace*{-8pt}

\bibliographystyle{Chicago}
\bibliography{refBib}


\end{document}